\documentclass[%
reprint,
superscriptaddress,
amsmath,amssymb,
aps,
pra,
]{revtex4-2}

\usepackage[dvipsnames]{xcolor}
\usepackage[utf8]{inputenc}
\usepackage[english]{babel}
\usepackage{amsmath}
\usepackage{graphicx}  
\usepackage{dcolumn} 
\usepackage{bm} 
\usepackage[hidelinks]{hyperref} 
\usepackage[detect-weight=true,separate-uncertainty=true]{siunitx}  
\usepackage{svg}
\usepackage{soul}
\usepackage{newfloat}
\usepackage[all]{hypcap}
\usepackage[capitalise,nameinlink]{cleveref}

\usepackage{xr}

\usepackage{lipsum}

\graphicspath{{figs/}}  

\definecolor{bluegray}{RGB}{40,180,160}
\definecolor{navygray}{RGB}{110,140,170}
\hypersetup{
citecolor={bluegray},
linkcolor={navygray},
urlcolor={bluegray},
}

\DeclareFloatingEnvironment[name={Figure S.\!\!}]{suppfigure}
\DeclareFloatingEnvironment[name={Table S.\!\!}]{supptable}

\crefname{suppfigure}{Fig. S.\!\!}{Figs. S.\!\!}
\Crefname{suppfigure}{Figure S.\!\!}{Figures S.\!\!}

\crefname{supptable}{Tab. S.\!\!}{Tabs. S.\!\!}
\Crefname{supptable}{Table S.\!\!}{Tables S.\!\!}

\newcommand{\bencomment}[1]{\emph{\color[rgb]{0.2,0,0.9}(BD: #1)}}

\newcommand{\benremove}[1]{{\color[rgb]{0.2,0,0.9}\sout{#1}}}
\renewcommand{\bencomment}[1]{{}}

\renewcommand{\benremove}[1]{{}}

\usepackage{epstopdf}
\usepackage{braket}
\usepackage{upgreek}
\usepackage{amsmath}
\usepackage{siunitx}
\usepackage{changes}
\usepackage{natbib}
\usepackage[title]{appendix}

\usepackage{multirow}

\begin{document}

\title{Superconducting circuits host two-level systems with lifetimes exceeding one second}


\author{Mathieu~F\'echant}
\thanks{The first two authors contributed equally.}
\affiliation{IQMT,~Karlsruhe~Institute~of~Technology,~76131~Karlsruhe,~Germany}

\author{Philipp~Lenhard}
\thanks{The first two authors contributed equally.}
\affiliation{IQMT,~Karlsruhe~Institute~of~Technology,~76131~Karlsruhe,~Germany}

\author{Nicolas~Gosling}
\affiliation{IQMT,~Karlsruhe~Institute~of~Technology,~76131~Karlsruhe,~Germany}

\author{Ritika~Dhundhwal}
\affiliation{IQMT,~Karlsruhe~Institute~of~Technology,~76131~Karlsruhe,~Germany}

\author{Denis~B\'en\^atre}
\affiliation{IQMT,~Karlsruhe~Institute~of~Technology,~76131~Karlsruhe,~Germany}

\author{Wolfgang~Wernsdorfer}
\affiliation{IQMT,~Karlsruhe~Institute~of~Technology,~76131~Karlsruhe,~Germany}
\affiliation{PHI,~Karlsruhe~Institute~of~Technology,~76131~Karlsruhe,~Germany}

\author{Thomas~Reisinger}
\affiliation{IQMT,~Karlsruhe~Institute~of~Technology,~76131~Karlsruhe,~Germany}

\author{Ioan~M.~Pop}
\email{ioan.pop@kit.edu}
\affiliation{IQMT,~Karlsruhe~Institute~of~Technology,~76131~Karlsruhe,~Germany}
\affiliation{PHI,~Karlsruhe~Institute~of~Technology,~76131~Karlsruhe,~Germany}
\affiliation{Physics~Institute~1,~Stuttgart~University,~70569~Stuttgart,~Germany}

\date{\today}


\begin{abstract}
Understanding the microscopic environment of superconducting qubits is a central challenge for quantum technologies. 
Parasitic two-level systems (TLSs) of various microscopic origins are ubiquitous and limit qubit performance through decoherence, frequency drifts, and leakage. 
Recently, long-lived TLSs with lifetimes exceeding those of state-of-the-art qubits have been shown to induce correlated qubit dynamics, with direct implications for quantum error correction.
Here, we perform wide-band quantum-jump correlation spectroscopy on  flux-tunable transmon qubits and identify microscopic degrees of freedom that outlive the qubit by orders of magnitude.
Pump–probe measurements at varying detunings from these objects reveal intrinsic lifetimes exceeding one second.
Their weak coupling to the qubit, in the kilohertz range, is consistent with positional defects described by the standard tunneling model, without excluding other microscopic origins such as magnetic spin impurities or trapped quasiparticles. 
More broadly, these results establish superconducting qubits as a sensitive probe of exceptionally long-lived microscopic degrees of freedom and open a route toward their controlled access and functionalization.
\end{abstract}

\maketitle 

With every gain in coherence, superconducting qubits have exposed new mechanisms by which they couple to their complex environment. Among the various sources of decoherence uncovered through the spectacular increase of qubit lifetimes over the years, a broad class of spurious two-level systems (TLSs) has been identified since two decades ago as one of the dominant contributions. They have been measured either as anti-crossings with the qubit spectrum, if the coupling is strong enough \cite{PhysRevLett.93.077003,Martinis2005Jul,Liu2024Oct,Burnett_2019}, or as drops in the qubit lifetime at specific frequencies, in the case of weak coupling \cite{Barends2013Aug}.
The problem of disentangling their microscopic origin between structural or chemical defects in dielectrics, trapped quasiparticles, surface spins, etc., which could all coexist, remains open \cite{de_Graaf_2020,Faoro2006,Un2022_decoherence,Caruzzo21,DelocalizedOxygen}. Nevertheless, systematic studies have resolved their distribution in frequency, presence in specific materials or positions in space, as well as their dependence on strain, electric field and qubit design \cite{Grabovskij2012Oct,Lisenfeld2015Feb,TempDependenceLisenfeld,Lisenfeld2019Nov,Bilmes2021Feb,Weeden2025May,Kim2025Sep,Reiss2025May,Yu2022Oct,Chen2023Oct,wang2025spectroscopycoherentcontroltwolevel}, enabling energy relaxation times to consistently improve \cite{joshi2026betatantalumtransmonqubits,  bland20252dtransmonslifetimescoherence, mscoherenceSomoroff}.  

Building on these developments, a new class of so-called long-lived TLSs, with lifetimes exceeding that of the qubit, have recently been discovered \cite{Spiecker2023Sep,Gosling2025,Zhuang2025Mar,Chang2025Nov,odeh2023nonmarkoviandynamicssuperconductingqubit}. When the qubit loses a photon to such a TLS, it can receive it back at a later time, breaking the fundamental assumption of a Markovian qubit environment. The resulting hidden temporal correlations in the qubit state could play a significant role in the dynamics of error codes and remain an active area of investigation \cite{PhysRevLett.96.050504,Mann2025Aug}.
While the presence of long-lived TLSs has now been established in several platforms, their intrinsic coherence remains an open question.

\begin{figure}[!t]
\includegraphics[width = \columnwidth]{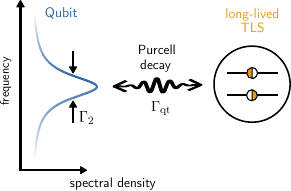}
\caption{\textbf{Schematic of the interaction between a qubit and a long-lived TLS}, in the case where they are each other's dominant loss mechanism. For low coupling $\mathrm{g}$ compared to the qubit and TLS decoherence rates,$\mathrm{\Gamma_2}$ and $\mathrm{(\Gamma_2)_{TLS}}$ respectively , vacuum Rabi oscillations are overdamped and the energy exchange between qubit and TLS is described by a Purcell rate $\mathrm{\Gamma_{qt}}$. }
\label{fig_schematic}
\end{figure}

\begin{figure*}
\hspace*{-0.5cm}
\includegraphics[width = 2.05\columnwidth]{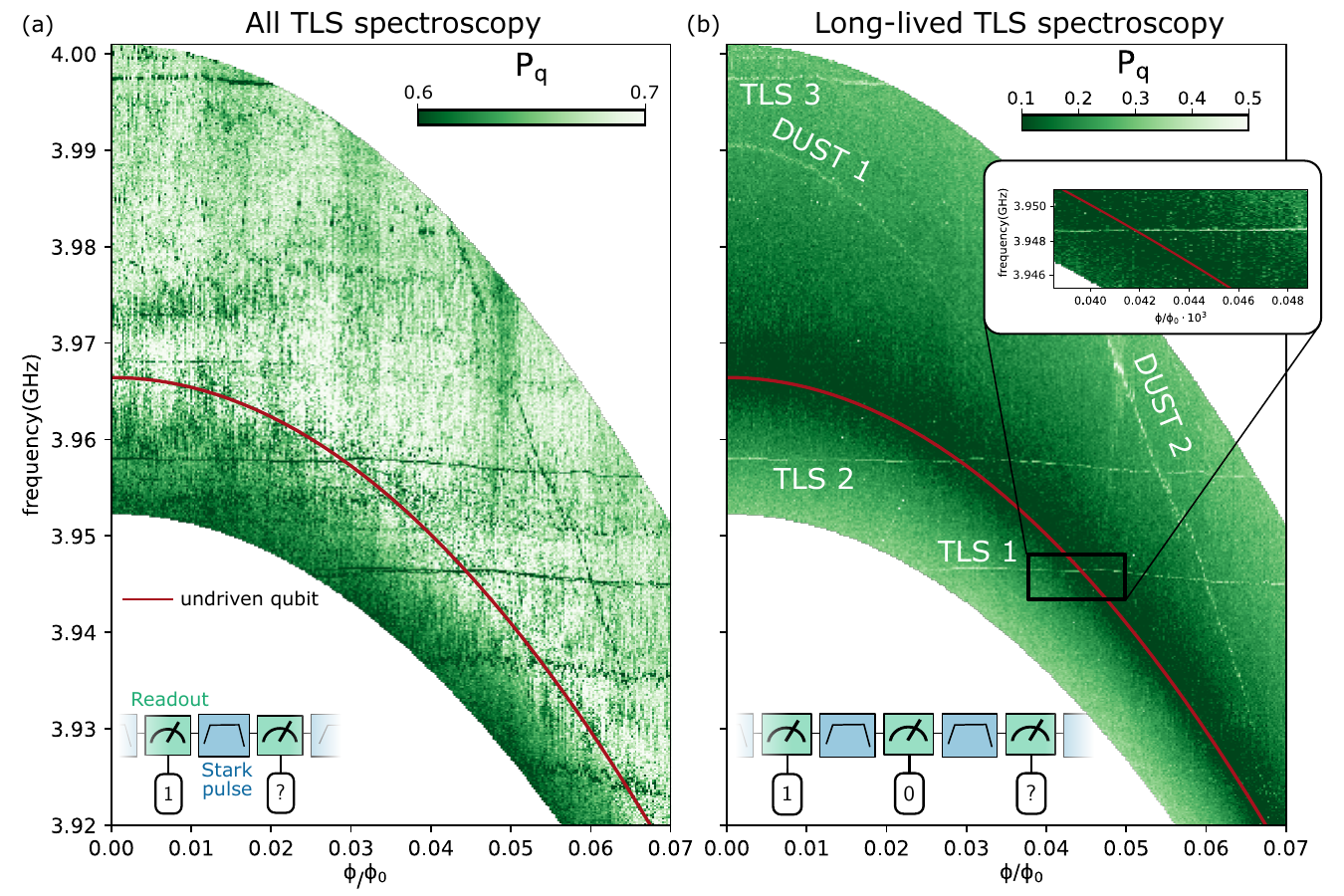}
\caption{\textbf{Quantum jump correlation spectroscopy performed over flux and Stark shift detuning.} Each point corresponds to a stroboscopic quantum jumps experiment with 1~$\mathrm{\mu s}$ measurement pulse at zero detuning and a 40~$\mathrm{\mu s}$ wait at a Stark shift detuned frequency. \textbf{(a)} Qubit population conditioned on measuring $\ket{1}$ in the previous step, as shown in the bottom left schematic. The range of qubit population from 0.6 to 0.7 corresponds to a range of $\mathrm{T_1}$ from 70 to 110~$\mathrm{\mu s}$. The red line depicts the frequency of the undriven flux tunable Transmon. \textbf{(b)} Qubit population conditioned on measuring a jump down, as shown in the bottom left schematic. This correlation spectroscopy shows photon back flow after a qubit jump down, revealing objects with lifetime comparable or longer than that of the qubit \cite{Gosling2025}. The 'TLS' labeled lines do not show flux dependence within the accuracy of the calibration. The 'DUST' labeled lines are flux dependent and likely originate from multi-photon processes involving the qubit, readout, TLS and Stark Shift pulse \cite{dai2026,jurgen_2026,Dumas2024}. The inset shows a zoom-in with increased calibration accuracy on TLS 1. }
\label{fig_transition_plot}
\end{figure*}

In this work, we isolate long-lived TLSs in the environment of a transmon qubit using quantum jump correlation spectroscopy \cite{Gosling2025}, and we introduce a pump-probe technique which allows us to measure their intrinsic relaxation time. We focus on three long-lived TLSs in two tantalum electrode transmon qubits with  standard $\mathrm{Al/AlO_x/Al}$ Josephson junctions.
When on resonance, the qubit constitutes the main decay channel for the TLS, via the Purcell effect (see Fig.~\ref{fig_schematic}), similar to a spin coupled to a resonator \cite{Bienfait2016Mar}.
We measure the intrinsic TLS lifetime as the saturation of the Purcell loss rate with qubit-TLS detuning, in the range of seconds. Moreover, we find a remarkably narrow Purcell linewidth in the range of 100 kHz, close to the qubit decoherence $\Gamma_2$, which also suggests a low TLS dephasing.

Our experimental platform consists of a tantalum transmon with an aluminium SQUID junction fabricated in a  two step shadow evaporation process~(see Supplementary~\ref{SUP_qubit_params}). We report on two devices with frequencies at the  zero-flux sweet spot of 3.9 GHz (Sample 1) and 3.1 GHz (Sample 2)  with Josephson energies $\mathrm{E_j}$ of 8.3~GHz and 14.42~GHz and charging energy $\mathrm{E_c}$ of 0.164~GHz and 0.147~GHz, respectively. The corresponding readout resonators for the two samples sit at 6.906~GHz and 7.493~GHz. The qubit state dependent dispersive shift $\mathrm{\chi}$ for both samples is approximately 2 MHz. The energy relaxation time $\mathrm{T_1}$ is in the range of 100 $\mathrm{\mu s}$ and shows commonly reported but poorly understood fluctuations depending on the cool down and the flux bias (see Supplementary \ref{SUP_qubit_params}) \cite{Klimov_2018,carroll2022dynamicssuperconductingqubitrelaxation,Burnett_2019}. Shedding light on the underlying processes behind these fluctuations constitutes one of the main motivations for this work.

To sweep the qubit frequency, we can either change the flux bias of the SQUID loop, which is slow compared to the qubit decay rate, or apply an ac Stark-shift pulse, which shifts the qubit frequency practically instantaneously. While the Stark shift pump can be detuned by as little as tens of MHz from the qubit \cite{Thorbeck2023Jun,schneider_ACstark18}, the optimal pump frequency can also be more than 1~GHz away from the qubit, where the qubit–resonator system is less susceptible to drive-induced unwanted state transitions (DUST) \cite{weidai26}. Depending on the specific qubit, resonator, environment configuration, as well as the target frequency, we choose one Stark shift strategy or the other, which allows us to shift the qubit frequency up or down by tens of MHz  (see Supplementary~\ref{SUP_Stark_shift}).

\begin{figure*}
\hspace*{-.4cm}
\includegraphics[width = 2.07\columnwidth]{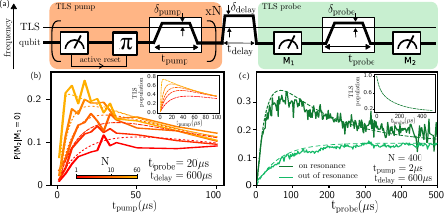}
\caption{ \textbf{TLS pump and probe.} \textbf{(a)} Schematic of the pump (orange highlight) and probe (green highlight) sequence. To pump the TLS, we actively reset the qubit after which we bring to a detuning $\delta_{pump}$ from the TLS for a time $\mathrm{t_{pump}}$. To fully polarize the TLS, we repeat the pumping sequence N times. After pumping, we idle at a detuning $\delta_{delay}$. To probe the TLS, we first measure the qubit (M1), after which we idle for $\mathrm{t_{probe}}$ at a detuning $\mathrm{\delta_{probe}}$ with the TLS, and finally we measure the population returned back to the qubit (M2). \textbf{(b)} Probe population after pumping sequences ranging from 0 to 100 $\mathrm{\mu s}$ and $N=1$ to $N=60$ repetitions. Dashed lines indicate the fit with Solomon equations with $\mathrm{\Gamma_{qt} = (70~\mu s)^{-1} }$ and $\mathrm{\Gamma_{q} = (50~\mu s)^{-1}}$ while the while the TLS population derived from this fit is shown in the inset. \textbf{(c)} Probe population after a saturation sequence on and off resonance, showing the TLS photon return over time. The dashed lines indicate the joint fit with the same parameters as in (b) with the corresponding TLS population shown in the inset. The higher TLS polarization reported in panel c compared to b is explained by the larger number of pumping cycles $\mathrm{N}$. For both (b) and (c) the detuning $\delta_{delay}$ is of multiple MHz.}
\label{fig_zoom}
\end{figure*}

\vspace{5mm}

As shown in Fig.~\ref{fig_transition_plot}a, we start by probing the environment of the qubit passively, i.e. without resonantly driving the qubit. Repeated 1~$\mathrm{\mu s}$ long single-shot readout is performed at constant qubit frequency with the Stark shift pump turned off. In order to map the qubit environment, in between the readout pulses, we detune the qubit using the Stark shift pump during 40~$\mathrm{\mu s}$ (cf. inset Fig.~\ref{fig_transition_plot}a). 
By averaging the qubit population $\mathrm{P_q}$ immediately following an excited state measurement, as a function of flux and Stark detuning, we observe signatures of various spurious modes in the qubit environment as excess decay. This broad scan shows different spectral line behaviors: narrow and stable, jittering  with varying broadness, which are both independent of flux and in agreement with numerous prior reports \cite{Grabovskij2012Oct,Lisenfeld2015Feb,TempDependenceLisenfeld,Lisenfeld2019Nov,Bilmes2021Feb,Weeden2025May,Kim2025Sep,Reiss2025May,Yu2022Oct,Chen2023Oct,wang2025spectroscopycoherentcontroltwolevel}, but also periodic flux dependent, which suggests DUST \cite{Dumas2024,weidai26,fechantPRL25}.

\begin{figure*}
\includegraphics[width = 2\columnwidth]{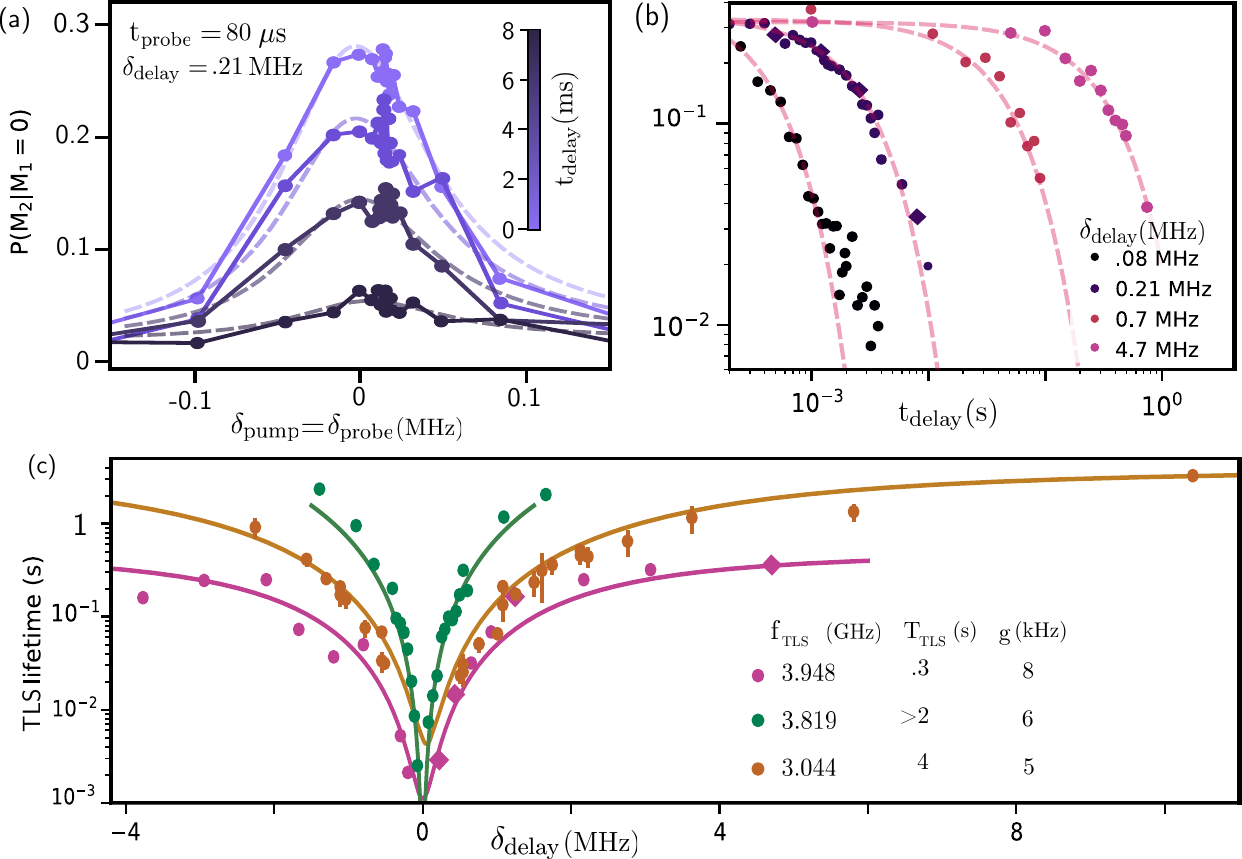}
\caption{\textbf{TLS Relaxation with qubit-TLS detuning.} \textbf{(a)} Return photon amplitude for increasing delay time showing the decay of the TLS photon. The dashed lines show the Lorentzian fit with width around .1 MHz, close to the qubit decoherence rate of $(15~\mathrm{\mu s})^{-1}$.  \textbf{(b)} Decay curves of the return population for varying qubit TLS detuning $\mathbf{\delta_{delay}}$. Dashed lines show the fitted exponential decay. The diamonds on the .21 MHz detuning curve indicate the points shown in (a)  \textbf{(c)} Measured TLS lifetimes for three different TLSs as a function of qubit-TLS detuning . The TLS at 3.948 GHz is TLS~1 in figure \ref{fig_transition_plot}. The solid lines show the fitted Purcell decay in which the width $\mathrm{\Gamma}$ of the Purcell formula is taken to be the qubit $\mathrm{T_2}$ fixed at 15, 50, and 7 $\mathrm{ \mu s}$. The diamonds on the magenta TLS indicate the detuning shown in (b). }
\label{fig_purcell}
\end{figure*}

To reveal the non-Markovian part of the environment, in Fig.~\ref{fig_transition_plot}b we deploy the correlation spectroscopy introduced in Ref.~\cite{Gosling2025}. Using the same dataset as in Fig.~\ref{fig_transition_plot}a, we average the qubit population after every downward jump (cf. schematics in the inset of Fig.~\ref{fig_transition_plot}b). The measurement of excess population indicates that the energy released during the jump down has excited a long-lived element of the environment, that subsequently relaxed back into the qubit. 
This method erases most of the lines visible in Fig.~\ref{fig_transition_plot}a, revealing three long-lived TLSs within the frequency span of 80 MHz. 
An example of the full time evolution after a quantum jump down with the qubit and TLS 1 in resonance is shown in Supplementary~\ref{Supp_in_resonance}, exhibiting similar non-Markovian behavior as in Ref.~\cite{Gosling2025}.

When resonant, long-lived TLSs lose energy predominantly through the qubit. To measure their intrinsic lifetime, we therefore perform a pump--probe experiment in which fast Stark shift detuning switches the interaction ON during pumping and probing and OFF during the waiting time (Fig.~\ref{fig_zoom}a). By resetting the qubit to $|1\rangle$ before each resonant interaction with the TLS, we progressively pump population into the TLS (see Fig~\ref{fig_zoom}b). We probe the photon amplitude in the TLS by conditioning the qubit in $|0\rangle$ (measure M1 in Fig.~\ref{fig_zoom}a) and waiting for the TLS excitation to populate the qubit (see Fig.~\ref{fig_zoom}c). We fit the evolution of qubit population both during pumping and probing using the Solomon equations \cite{Spiecker_2024}, which allows us to infer the TLS population during the experiment.
This fit reveals that by repeating the pumping cycle $N$-times we can polarize the TLS, as shown in insets of Fig.~\ref{fig_zoom}b,c.
While in our case the preparation acts as a Szilard engine operating on a single TLS, if the qubit decoherence rate would be smaller than the TLS--qubit coupling, this process would become deterministic, requiring only a single swap operation. In the case of the qubit and TLSs reported in this work, we are within an order of magnitude from accessing this strong coupling regime.

To reveal the intrinsic TLS lifetime, we use the pump-probe protocol of Fig.~\ref{fig_zoom} to perform TLS spectroscopy on TLS 1, for several delay intervals $\mathrm{t_{delay}}$ between the pump and probe. As shown in Fig.~\ref{fig_purcell}a, even for a modest qubit-TLS detuning $\mathrm{\delta_{delay}}$ of 210 kHz, the TLS linewidth remains visible for several milliseconds. With higher detuning we uncover a dramatic increase of TLS lifetime, as shown by the measured decay curves in Fig.~\ref{fig_purcell}b. In Fig.~\ref{fig_purcell}c we summarize the measured intrinsic TLS relaxation times as a function of detuning for TLS 1 and two other subsequently measured TLSs in different cooldowns (see Supplementary~\ref{SUP_otherstark_shift}). We fit the Purcell-limited decay rate of each TLS as a function of detuning using
\begin{equation}
    \Gamma_\text{TLS}(\delta) = \Gamma_\text{TLS}^{0} + \frac{2 g^2 \Gamma}{\Gamma^2 + \mathrm{\delta_{delay}}^2},
    \label{eq:purcell}
\end{equation}
where $g$ is the TLS--qubit coupling strength, $\mathrm{\Gamma = \Gamma_2 + (\Gamma_2)_{TLS}}$ is taken to be equal to the qubit linewidth $\Gamma_2$, and $\Gamma_\text{TLS}^{0}$ is the detuning-independent intrinsic decay rate of the TLS~\cite{Bienfait2016Mar}. Remarkably, we find two TLSs with intrinsic lifetimes exceeding one second. Moreover, the  100~kHz TLS linewidth (cf Fig.~\ref{fig_purcell}a) indicates the qubit as the dominant dephasing mechanism and suggests not only a long-lived, but also a remarkably coherent TLS family.

We can use the measurements reported in Fig.~\ref{fig_purcell}c, which go orders of magnitude beyond previously reported intrinsic lifetimes, to provide new constraints for microscopic TLS models. Using the generic Hamiltonian of a positional defect tunneling between two wells, i.e. the so-called standard tunneling model~\cite{phillips_TLS,Anderson1972}, we can write the excitation energy as $\mathrm{\epsilon = \sqrt{\Delta^2 + \Lambda^2}}$ where $\mathrm{\Delta}$ is the potential asymmetry and $\mathrm{\Lambda}$ the tunnel splitting. 
Given that both $\Delta$ and $\Lambda$ control the energy splitting independently, at any given TLS frequency $\mathrm{\epsilon/h}$, it is possible to find an object with an arbitrary low tunneling $\mathrm{\Lambda}$, potentially implementing a long-lived TLS. 
The expected maximal coupling to the qubit scales as $\mathrm{\Lambda}$, while the effective TLS lifetime is limited by phonon losses, scaling as as $\mathrm{\Lambda}^2$ via Fermi's golden rule~\cite{phillips_TLS}. For TLSs with coupling g in the MHz, reported intrinsic relaxation times are about 1~$\mathrm{\mu s}$ \cite{Shalibo_2010}. Extrapolating to kHz coupling, we would expect intrinsic decay times in the seconds, consistent with our measurement  (cf Fig.~\ref{fig_purcell}c). This indicates that the standard tunneling model for positional defects, while not determining the microscopic nature of the long-lived TLSs, can account for the long lifetimes. 

Assuming defects tunnel between uniformly distributed positions in space, the number of TLSs is highly skewed towards low tunneling, with correspondingly low coupling and long lifetimes. As material engineering and surface treatment successfully decreases the amount of TLSs, statistically it becomes increasingly likely to find frequency ranges  where the qubit interaction is dominated by long-lived TLSs. The situation is similar to a telescope with increasingly better focus and narrow field, picking out increasingly dimmer stars in regions previously assumed to be dark. Therefore, within this model, we can expect that it will be commonplace for future higher quality qubits to be mostly coupled to long-lived TLSs. In this scenario, each order of magnitude increase in qubit lifetimes unlocks two orders of magnitude higher coherence in the TLS subspace.



In summary, we have shown that a superconducting circuit's environment hosts TLSs with intrinsic lifetimes extending to seconds, orders of magnitude longer than that of the qubit. These long-lived TLSs constitute a coherent but unintended memory channel for the qubit, while their weak coupling makes them difficult to detect with conventional spectroscopy. While the scaling of the measured coupling and intrinsic relaxation time is consistent with tunneling positional defects, pinpointing their microscopic origin will require further work, such as testing for an electric and magnetic field  dependence and measuring their dephasing directly. Our results establish a quantitative framework for characterizing TLS-induced decoherence, and also suggest that long-lived TLSs could be harnessed as quantum memory elements in superconducting circuits.

\section*{Acknowledgments}

We thank Sumeru Hazra, Martin Spiecker, Wulf Wulfhekel, Jürgen Lisenfeld, Hannes Rotzinger, Lara Faoro and Lev Ioffe for fruitful discussions.
We thank N. Zapata and M. Field for helping with the readout amplification chain.
We thank L. Radtke and S. Diewald for technical assistance.
This work was financed by the German Ministry of Education and Research (BMBF) within project QSolid (FKZ:13N16151).
D.B.~acknowledges funding from the Horizon Europe program via Project No.~101113946 OpenSuperQPlus100.
N.G.~acknowledges funding from the German Ministry of Education and Research (BMBF) within project GEQCOS (FKZ:~13N15683).
We acknowledge funding from Google via their academic support initiative.
Facilities use was supported by the KIT Nanostructure Service Laboratory. 
We acknowledge the measurement software framework qKit. 
During the writing of this manuscript, we became aware of recently uploaded related results~\cite{2026quantumenvironmentafterglowbroadband}.

\bibliography{TLS_references}

@article{Spiecker_2024,
   title={Solomon equations for qubit and two-level systems: Insights into non-Poissonian quantum jumps},
   volume={109},
   ISSN={2469-9934},
   url={http://dx.doi.org/10.1103/PhysRevA.109.052218},
   DOI={10.1103/physreva.109.052218},
   number={5},
   journal={Physical Review A},
   publisher={American Physical Society (APS)},
   author={Spiecker, Martin and Pavlov, Andrei I. and Shnirman, Alexander and Pop, Ioan M.},
   year={2024},
   month=May }

@article{Maleeva_2018,
   title={Circuit quantum electrodynamics of granular aluminum resonators},
   volume={9},
   ISSN={2041-1723},
   url={http://dx.doi.org/10.1038/s41467-018-06386-9},
   DOI={10.1038/s41467-018-06386-9},
   number={1},
   journal={Nature Communications},
   publisher={Springer Science and Business Media LLC},
   author={Maleeva, N. and Grünhaupt, L. and Klein, T. and Levy-Bertrand, F. and Dupre, O. and Calvo, M. and Valenti, F. and Winkel, P. and Friedrich, F. and Wernsdorfer, W. and Ustinov, A. V. and Rotzinger, H. and Monfardini, A. and Fistul, M. V. and Pop, I. M.},
   year={2018},
   month=Sept }

@article{Chen2023Oct,
   title={Phonon engineering of atomic-scale defects in superconducting quantum circuits},
   volume={10},
   ISSN={2375-2548},
   url={http://dx.doi.org/10.1126/sciadv.ado6240},
   DOI={10.1126/sciadv.ado6240},
   pages={37},
   journal={Science Advances},
   publisher={American Association for the Advancement of Science (AAAS)},
   author={Chen, Mo and Owens, John Clai and Putterman, Harald and Schäfer, Max and Painter, Oskar},
   year={2024},
   month=Sept }

@article{Zhuang2025Mar,
    author = {Zhuang, Ze-Tong and Rosenstock, Dario and Liu, Bao-Jie and Somoroff, Aaron and Manucharyan, Vladimir E. and Wang, Chen},
      title   = {Non-Markovian relaxation spectroscopy of fluxonium qubits},
    journal = {Nature Communications},
    year    = {2026},
    doi     = {10.1038/s41467-026-69910-2},
    url     = {https://www.nature.com/articles/s41467-026-69910-2}
}

@article{PhysRevLett.96.050504,
  title = {Fault-Tolerant Quantum Computation with Long-Range Correlated Noise},
  author = {Aharonov, Dorit and Kitaev, Alexei and Preskill, John},
  journal = {Phys. Rev. Lett.},
  volume = {96},
  issue = {5},
  pages = {050504},
  numpages = {4},
  year = {2006},
  month = {Feb},
  publisher = {American Physical Society},
  doi = {10.1103/PhysRevLett.96.050504},
  url = {https://link.aps.org/doi/10.1103/PhysRevLett.96.050504}
}

@article{jurgen_2026,
   title={Readout failures in superconducting qubits due to TLS-defects in tunnel junctions},
   ISSN={2643-1564},
  author={Lisenfeld, Jürgen and Händel, Alexander K. and Bilmes, Alexander and Ustinov, Alexey V.},
   url={http://dx.doi.org/10.1103/ypm5-lh8d},
   DOI={10.1103/ypm5-lh8d},
   journal={Physical Review Research},
   publisher={American Physical Society (APS)},
   author={Anonymous},
   year={2026},
   month=Sept }

@article{fechantPRL25,
  title = {Offset Charge Dependence of Measurement-Induced Transitions in Transmons},
  author = {F{\'e}chant, Mathieu and Dumas, Marie Fr{\'e}d{\'e}rique and B{\'e}n{\^a}tre, Denis and Gosling, Nicolas and Lenhard, Philipp and Spiecker, Martin and Geisert, Simon and Ihssen, S{\"o}ren and Wernsdorfer, Wolfgang and D'Anjou, Benjamin and Blais, Alexandre and Pop, Ioan M.},
  journal = {Phys. Rev. Lett.},
  volume = {135},
  issue = {18},
  pages = {180603},
  numpages = {7},
  year = {2025},
  month = {Oct},
  publisher = {American Physical Society},
  doi = {10.1103/yljv-b4kj},
  url = {https://link.aps.org/doi/10.1103/yljv-b4kj}
}

@article{Dumas2024,
  title = {Measurement-Induced Transmon Ionization},
  author = {Dumas, Marie Fr\'ed\'erique and Groleau-Par\'e, Benjamin and McDonald, Alexander and Mu\~noz-Arias, Manuel H. and Lled\'o, Crist\'obal and D'Anjou, Benjamin and Blais, Alexandre},
  journal = {Phys. Rev. X},
  volume = {14},
  issue = {4},
  pages = {041023},
  numpages = {31},
  year = {2024},
  month = {Oct},
  publisher = {American Physical Society},
  doi = {10.1103/PhysRevX.14.041023},
  url = {https://link.aps.org/doi/10.1103/PhysRevX.14.041023}
}

@article{weidai26,
  title = {Characterization of Drive-Induced Unwanted State Transitions in Superconducting Circuits},
  author = {Dai, W. and Hazra, S. and Weiss, D. K. and Kurilovich, P. D. and Connolly, T. and Babla, H. K. and Singh, S. and Joshi, V. R. and Ding, A. Z. and Parakh, P. D. and Venkatraman, J. and Xiao, X. and Frunzio, L. and Devoret, M. H.},
  journal = {Phys. Rev. X},
  volume = {16},
  issue = {1},
  pages = {011011},
  numpages = {26},
  year = {2026},
  month = {Jan},
  publisher = {American Physical Society},
  doi = {10.1103/zdpg-mhpc},
  url = {https://link.aps.org/doi/10.1103/zdpg-mhpc}
}

@misc{2026quantumenvironmentafterglowbroadband,
      title={Quantum environment afterglow from broadband excitation spectroscopy in superconducting qubits}, 
      author={J. R. Guimarães and Y. Liu and Y. Gao and Y. Haddad and A. Galicia and D. A. Volkov and J. T. Schmieder and H. Bhardwaj and M. Neis and J. Cereijo and M. Jerger and P. A. Bushev and R. Barends},
      year={2026},
      eprint={2609.31280},
      archivePrefix={arXiv},
      
      url={https://arxiv.org/abs/2609.31280}, 
}

@article{schneider_ACstark18,
  title = {Local sensing with the multilevel ac Stark effect},
  author = {Schneider, Andre and Braum\"uller, Jochen and Guo, Lingzhen and Stehle, Patrizia and Rotzinger, Hannes and Marthaler, Michael and Ustinov, Alexey V. and Weides, Martin},
  journal = {Phys. Rev. A},
  volume = {97},
  issue = {6},
  pages = {062334},
  numpages = {8},
  year = {2018},
  month = {Jun},
  publisher = {American Physical Society},
  doi = {10.1103/PhysRevA.97.062334},
  url = {https://link.aps.org/doi/10.1103/PhysRevA.97.062334}
}

@article{dai2026,
  title = {Characterization of Drive-Induced Unwanted State Transitions in Superconducting Circuits},
  author = {Dai, W. and Hazra, S. and Weiss, D. K. and Kurilovich, P. D. and Connolly, T. and Babla, H. K. and Singh, S. and Joshi, V. R. and Ding, A. Z. and Parakh, P. D. and Venkatraman, J. and Xiao, X. and Frunzio, L. and Devoret, M. H.},
  journal = {Phys. Rev. X},
  volume = {16},
  issue = {1},
  pages = {011011},
  numpages = {26},
  year = {2026},
  month = {Jan},
  publisher = {American Physical Society},
  doi = {10.1103/zdpg-mhpc},
  url = {https://link.aps.org/doi/10.1103/zdpg-mhpc}
}

@article{Anderson1972,
  author  = {P. W. Anderson and B. I. Halperin and C. M. Varma},
  title   = {Anomalous Low-Temperature Thermal Properties of Glasses and Spin Glasses},
  journal = {Philosophical Magazine},
  volume  = {25},
  number  = {1},
  pages   = {1--9},
  year    = {1972},
  doi     = {10.1080/14786437208229210}
}

@article{Burnett_2019,
   title={Decoherence benchmarking of superconducting qubits},
   volume={5},
   ISSN={2056-6387},
   url={http://dx.doi.org/10.1038/s41534-019-0168-5},
   DOI={10.1038/s41534-019-0168-5},
   number={1},
   journal={npj Quantum Information},
   publisher={Springer Science and Business Media LLC},
   author={Burnett, Jonathan J. and Bengtsson, Andreas and Scigliuzzo, Marco and Niepce, David and Kudra, Marina and Delsing, Per and Bylander, Jonas},
   year={2019},
   month=June }

@misc{carroll2022dynamicssuperconductingqubitrelaxation,
      title={Dynamics of superconducting qubit relaxation times}, 
      author={Malcolm Carroll and Sami Rosenblatt and Petar Jurcevic and Isaac Lauer and Abhinav Kandala},
      year={2022},
      eprint={2105.15201},
      archivePrefix={arXiv},
      url={https://arxiv.org/abs/2105.15201}, 
}

@article{Klimov_2018,
   title={Fluctuations of Energy-Relaxation Times in Superconducting Qubits},
   volume={121},
   ISSN={1079-7114},
   url={http://dx.doi.org/10.1103/PhysRevLett.121.090502},
   DOI={10.1103/physrevlett.121.090502},
   number={9},
   journal={Phys. Rev. Lett.},
   publisher={American Physical Society (APS)},
   author={Klimov, P. V. and Kelly, J. and Chen, Z. and Neeley, M. and Megrant, A. and Burkett, B. and Barends, R. and Arya, K. and Chiaro, B. and Chen, Yu and Dunsworth, A. and Fowler, A. and Foxen, B. and Gidney, C. and Giustina, M. and Graff, R. and Huang, T. and Jeffrey, E. and Lucero, Erik and Mutus, J. Y. and Naaman, O. and Neill, C. and Quintana, C. and Roushan, P. and Sank, Daniel and Vainsencher, A. and Wenner, J. and White, T. C. and Boixo, S. and Babbush, R. and Smelyanskiy, V. N. and Neven, H. and Martinis, John M.},
   year={2018},
   month=Aug }

@misc{Gosling2025,
      title={Probing the memory of a superconducting qubit environment}, 
      author={Nicolas Gosling and Denis Bénâtre and Nicolas Zapata and Paul Kugler and Mitchell Field and Sumeru Hazra and Simon Günzler and Thomas Reisinger and Martin Spiecker and Mathieu Féchant and Ioan M. Pop},
      year={2026},
      eprint={2603.11889},
      archivePrefix={arXiv},
      url={https://arxiv.org/abs/2603.11889}, 
}

@article{Spiecker2023Sep,
	author = {Spiecker, Martin and Paluch, Patrick and Gosling, Nicolas and Drucker, Niv and Matityahu, Shlomi and Gusenkova, Daria and G{\ifmmode\ddot{u}\else\"{u}\fi}nzler, Simon and Rieger, Dennis and Takmakov, Ivan and Valenti, Francesco and Winkel, Patrick and Gebauer, Richard and Sander, Oliver and Catelani, Gianluigi and Shnirman, Alexander and Ustinov, Alexey V. and Wernsdorfer, Wolfgang and Cohen, Yonatan and Pop, Ioan M.},
	title = {{Two-level system hyperpolarization using a quantum Szilard engine}},
	journal = {Nat. Phys.},
	volume = {19},
	pages = {1320--1325},
	year = {2023},
	month = sep,
	issn = {1745-2481},
	publisher = {Nature Publishing Group},
	doi = {10.1038/s41567-023-02082-8}
}

@article{Thorbeck2023Jun,
	author = {Thorbeck, Ted and Eddins, Andrew and Lauer, Isaac and McClure, Douglas T. and Carroll, Malcolm},
	title = {{Two-Level-System Dynamics in a Superconducting Qubit Due to Background Ionizing Radiation}},
	journal = {PRX Quantum},
	volume = {4},
	number = {2},
	pages = {020356},
	year = {2023},
	month = jun,
	publisher = {American Physical Society},
	doi = {10.1103/PRXQuantum.4.020356}
}

@article{Martinis2005Jul,
  title = {Decoherence in Josephson Qubits from Dielectric Loss},
  author = {Martinis, John M. and Cooper, K. B. and McDermott, R. and Steffen, Matthias and Ansmann, Markus and Osborn, K. D. and Cicak, K. and Oh, Seongshik and Pappas, D. P. and Simmonds, R. W. and Yu, Clare C.},
  journal = {Phys. Rev. Lett.},
  volume = {95},
  issue = {21},
  pages = {210503},
  numpages = {4},
  year = {2005},
  month = {Nov},
  publisher = {American Physical Society},
  doi = {10.1103/PhysRevLett.95.210503},
  url = {https://link.aps.org/doi/10.1103/PhysRevLett.95.210503}
}

@article{Reiss2025May,
	author = {Reiss, Yigal and Schechter, Moshe},
	title = {{Thermal cycling: Evidence for a generalized tunneling model and a tool to distinguish noise sources in quantum circuits}},
	journal = {Phys. Rev. B},
	volume = {111},
	number = {20},
	pages = {L201403},
	year = {2025},
	month = may,
	publisher = {American Physical Society},
	doi = {10.1103/PhysRevB.111.L201403}
}

@article{Barends2013Aug,
	author = {Barends, R. and Kelly, J. and Megrant, A. and Sank, D. and Jeffrey, E. and Chen, Y. and Yin, Y. and Chiaro, B. and Mutus, J. and Neill, C. and O{'}Malley, P. and Roushan, P. and Wenner, J. and White, T. C. and Cleland, A. N. and Martinis, John M.},
	title = {{Coherent Josephson Qubit Suitable for Scalable Quantum Integrated Circuits}},
	journal = {Phys. Rev. Lett.},
	volume = {111},
	number = {8},
	pages = {080502},
	year = {2013},
	month = aug,
	publisher = {American Physical Society},
	doi = {10.1103/PhysRevLett.111.080502}
}

@article{Grabovskij2012Oct,
	author = {Grabovskij, Grigorij J. and Peichl, Torben and Lisenfeld, J{\ifmmode\ddot{u}\else\"{u}\fi}rgen and Weiss, Georg and Ustinov, Alexey V.},
	title = {{Strain Tuning of Individual Atomic Tunneling Systems Detected by a Superconducting Qubit}},
	journal = {Science},
	volume = {338},
	number = {6104},
	pages = {232--234},
	year = {2012},
	month = oct,
	issn = {0036-8075},
	publisher = {American Association for the Advancement of Science},
	doi = {10.1126/science.1226487}
}

@article{Lisenfeld2019Nov,
	author = {Lisenfeld, J{\ifmmode\ddot{u}\else\"{u}\fi}rgen and Bilmes, Alexander and Megrant, Anthony and Barends, Rami and Kelly, Julian and Klimov, Paul and Weiss, Georg and Martinis, John M. and Ustinov, Alexey V.},
	title = {{Electric field spectroscopy of material defects in transmon qubits}},
	journal = {npj Quantum Inf.},
	volume = {5},
	number = {105},
	pages = {1--6},
	year = {2019},
	month = nov,
	issn = {2056-6387},
	publisher = {Nature Publishing Group},
	doi = {10.1038/s41534-019-0224-1}
}

@article{Weeden2025May,
	author = {Weeden, S. and Harrison, D. C. and Patel, S. and Snyder, M. and Blackwell, E. J. and Spahn, G. and Abdullah, S. and Takeda, Y. and Plourde, B. L. T. and Martinis, J. M. and McDermott, R.},
	title = {{Statistics of Strongly Coupled Defects in Superconducting Qubits}},
	journal = {arXiv},
	year = {2025},
	month = {may},
	doi = {10.48550/arXiv.2506.00193}
}

@article{Liu2024Oct,
	author = {Liu, Bao-Jie and Wang, Ying-Ying and Sheffer, Tal and Wang, Chen},
	title = {{Observation of Discrete Charge States of a Coherent Two-Level System in a Superconducting Qubit}},
	journal = {Phys. Rev. Lett.},
	volume = {133},
	number = {16},
	pages = {160602},
	year = {2024},
	month = oct,
	publisher = {American Physical Society},
	doi = {10.1103/PhysRevLett.133.160602}
}

@article{TempDependenceLisenfeld,
  title = {Measuring the Temperature Dependence of Individual Two-Level Systems by Direct Coherent Control},
  author = {Lisenfeld, J. and M\"uller, C. and Cole, J. H. and Bushev, P. and Lukashenko, A. and Shnirman, A. and Ustinov, A. V.},
  journal = {Phys. Rev. Lett.},
  volume = {105},
  issue = {23},
  pages = {230504},
  numpages = {4},
  year = {2010},
  month = {Dec},
  publisher = {American Physical Society},
  doi = {10.1103/PhysRevLett.105.230504},
  url = {https://link.aps.org/doi/10.1103/PhysRevLett.105.230504}
}

@article{DelocalizedOxygen,
  title = {Delocalized Oxygen as the Origin of Two-Level Defects in Josephson Junctions},
  author = {DuBois, Timothy C. and Per, Manolo C. and Russo, Salvy P. and Cole, Jared H.},
  journal = {Phys. Rev. Lett.},
  volume = {110},
  issue = {7},
  pages = {077002},
  numpages = {5},
  year = {2013},
  month = {Feb},
  publisher = {American Physical Society},
  doi = {10.1103/PhysRevLett.110.077002},
  url = {https://link.aps.org/doi/10.1103/PhysRevLett.110.077002}
}

@article{Lisenfeld2015Feb,
	author = {Lisenfeld, J{\ifmmode\ddot{u}\else\"{u}\fi}rgen and Grabovskij, Grigorij J. and M{\ifmmode\ddot{u}\else\"{u}\fi}ller, Clemens and Cole, Jared H. and Weiss, Georg and Ustinov, Alexey V.},
	title = {{Observation of directly interacting coherent two-level systems in an amorphous material}},
	journal = {Nat. Commun.},
	volume = {6},
	number = {6182},
	pages = {6182},
	year = {2015},
	month = feb,
	issn = {2041-1723},
	publisher = {Nature Publishing Group},
	doi = {10.1038/ncomms7182}
}

@article{Faoro2006,
  title = {Quantum Two Level Systems and Kondo-Like Traps as Possible Sources of Decoherence in Superconducting Qubits},
  author = {Faoro, Lara and Ioffe, Lev B.},
  journal = {Phys. Rev. Lett.},
  volume = {96},
  issue = {4},
  pages = {047001},
  numpages = {4},
  year = {2006},
  month = {Jan},
  publisher = {American Physical Society},
  doi = {10.1103/PhysRevLett.96.047001},
  url = {https://link.aps.org/doi/10.1103/PhysRevLett.96.047001}
}

@article{de_Graaf_2020,
   title={Two-level systems in superconducting quantum devices due to trapped quasiparticles},
   volume={6},
   ISSN={2375-2548},
   url={http://dx.doi.org/10.1126/sciadv.abc5055},
   DOI={10.1126/sciadv.abc5055},
   pages={51},
   journal={Science Advances},
   publisher={American Association for the Advancement of Science (AAAS)},
   author={de Graaf, S. E. and Faoro, L. and Ioffe, L. B. and Mahashabde, S. and Burnett, J. J. and Lindström, T. and Kubatkin, S. E. and Danilov, A. V. and Tzalenchuk, A. Ya.},
   year={2020},
   month=Dec }

@article{mscoherenceSomoroff,
  title = {Millisecond Coherence in a Superconducting Qubit},
  author = {Somoroff, Aaron and Ficheux, Quentin and Mencia, Raymond A. and Xiong, Haonan and Kuzmin, Roman and Manucharyan, Vladimir E.},
  journal = {Phys. Rev. Lett.},
  volume = {130},
  issue = {26},
  pages = {267001},
  numpages = {6},
  year = {2023},
  month = {Jun},
  publisher = {American Physical Society},
  doi = {10.1103/PhysRevLett.130.267001},
  url = {https://link.aps.org/doi/10.1103/PhysRevLett.130.267001}
}

@article{bland20252dtransmonslifetimescoherence,
	author = {Bland, Matthew P. and Bahrami, Faranak and Martinez, Jeronimo G. C. and Prestegaard, Paal H. and Smitham, Basil M. and Joshi, Atharv and Hedrick, Elizabeth and Kumar, Shashwat and Yang, Ambrose and Pakpour-Tabrizi, Alexander C. and Jindal, Apoorv and Chang, Ray D. and Cheng, Guangming and Yao, Nan and Cava, Robert J. and de Leon, Nathalie P. and Houck, Andrew A.},
	title = {{Millisecond lifetimes and coherence times in 2D transmon qubits}},
	journal = {Nature},
	volume = {647},
	pages = {343--348},
	year = {2025},
	month = nov,
	issn = {1476-4687},
	publisher = {Nature Publishing Group},
	doi = {10.1038/s41586-025-09687-4}
}

@misc{joshi2026betatantalumtransmonqubits,
      title={Beta Tantalum Transmon Qubits with Quality Factors Approaching 10 Million}, 
      author={Atharv Joshi and Apoorv Jindal and Paal H. Prestegaard and Faranak Bahrami and Elizabeth Hedrick and Matthew P. Bland and Tunmay Gerg and Guangming Cheng and Nan Yao and Robert J. Cava and Andrew A. Houck and Nathalie P. de Leon},
      year={2026},
      eprint={2603.13174},
      archivePrefix={arXiv},
      url={https://arxiv.org/abs/2603.13174}, 
}

@article{Bilmes2021Feb,
	author = {Bilmes, Alexander and Volosheniuk, Serhii and Brehm, Jan David and Ustinov, Alexey V. and Lisenfeld, J{\ifmmode\ddot{u}\else\"{u}\fi}rgen},
	title = {{Quantum sensors for microscopic tunneling systems}},
	journal = {npj Quantum Inf.},
	volume = {7},
	number = {27},
	pages = {27},
	year = {2021},
	month = feb,
	issn = {2056-6387},
	publisher = {Nature Publishing Group},
	doi = {10.1038/s41534-020-00359-x}
}

@article{Caruzzo21,
  title = {Distribution of two-level system couplings to strain and electric fields in glasses at low temperatures},
  author = {Carruzzo, Herve M. and Bilmes, Alexander and Lisenfeld, J\"urgen and Yu, Zheng and Wang, Bu and Wan, Zhongyi and Schmidt, J. R. and Yu, Clare C.},
  journal = {Phys. Rev. B},
  volume = {104},
  issue = {13},
  pages = {134203},
  numpages = {10},
  year = {2021},
  month = {Oct},
  publisher = {American Physical Society},
  doi = {10.1103/PhysRevB.104.134203},
  url = {https://link.aps.org/doi/10.1103/PhysRevB.104.134203}
}

@article{Un2022_decoherence,
  author = {Un, Sun and de Graaf, Sebastian E. and Bertet, Patrice and Kubatkin, Sergey E. and Danilov, Andrey V.},
  title = {On the nature of decoherence in quantum circuits: Revealing the structural motif of the surface radicals in $\alpha$-{Al}$_2${O}$_3$},
  journal = {Science Advances},
  volume = {8},
  pages = {14},
  year = {2022},
  doi = {10.1126/sciadv.abm6169},
  url = {https://www.science.org/doi/10.1126/sciadv.abm6169},
}

@article{Chang2025Nov,
	author = {Chang, Tikai and Holzman, Itamar and Lim, Shao Qi and Holmes, Danielle and Johnson, Brett C. and Jamieson, David N. and Stern, Michael},
	title = {{Strong coupling of a superconducting flux qubit to single bismuth donors}},
	journal = {Nat. Commun.},
	volume = {16},
	number = {9832},
	pages = {9832},
	year = {2025},
	month = nov,
	issn = {2041-1723},
	publisher = {Nature Publishing Group},
	doi = {10.1038/s41467-025-64757-5}
}

@misc{odeh2023nonmarkoviandynamicssuperconductingqubit,
      title={Non-Markovian dynamics of a superconducting qubit in a phononic bandgap}, 
      author={Mutasem Odeh and Kadircan Godeneli and Eric Li and Rohin Tangirala and Haoxin Zhou and Xueyue Zhang and Zi-Huai Zhang and Alp Sipahigil},
      year={2023},
      eprint={2312.01031},
      archivePrefix={arXiv},
      primaryClass={quant-ph},
      url={https://arxiv.org/abs/2312.01031}, 
}

@article{PhysRevLett.93.077003,
  title = {Decoherence in Josephson Phase Qubits from Junction Resonators},
  author = {Simmonds, R. W. and Lang, K. M. and Hite, D. A. and Nam, S. and Pappas, D. P. and Martinis, John M.},
  journal = {Phys. Rev. Lett.},
  volume = {93},
  issue = {7},
  pages = {077003},
  numpages = {4},
  year = {2004},
  month = {Aug},
  publisher = {American Physical Society},
  doi = {10.1103/PhysRevLett.93.077003},
  url = {https://link.aps.org/doi/10.1103/PhysRevLett.93.077003}
}

@article{phillips_TLS,
    title={Two-level states in glasses},
    volume={50},
    url={https://iopscience.iop.org/article/10.1088/0034-4885/50/12/003},
    DOI={10.1088/0034-4885/50/12/003},
    pages={1657},
    journal={Rep. Prog. Phys.},
    author={W A Phillips},
    year={1987},
}

@article{Shalibo_2010,
   title={Lifetime and Coherence of Two-Level Defects in a Josephson Junction},
   volume={105},
   ISSN={1079-7114},
   url={http://dx.doi.org/10.1103/PhysRevLett.105.177001},
   DOI={10.1103/physrevlett.105.177001},
   pages={17},
   journal={Phys. Rev. Lett.},
   publisher={American Physical Society (APS)},
   author={Shalibo, Yoni and Rofe, Ya’ara and Shwa, David and Zeides, Felix and Neeley, Matthew and Martinis, John M. and Katz, Nadav},
   year={2010},
   month=Oct }

@article{wang2025spectroscopycoherentcontroltwolevel,
    author={Wang, Qianxu and Salcedo-Gallo, Juan S and Bedkihal, Salil and Xia, Tian and Olszewski, Maciej W and Fatemi, Valla and Fitzpatrick, Mattias},
   title={Spectroscopy and coherent control of two-level system defect ensembles using a broadband 3D waveguide},
   volume={5},
   journal={Materials for Quantum Technology},
   ISSN={2633-4356},
   url={http://dx.doi.org/10.1088/2633-4356/ae2b43},
   DOI={10.1088/2633-4356/ae2b43},
   number={4},
   publisher={IOP Publishing},
   
   year={2025},
   month={Dec}, 
   pages={045201} }

@article{Kim2025Sep,
	author = {Kim, Richard H. J. and Haeuser, Samuel J. and Park, Joong-Mok and Chan, Randall K. and Oh, Jin-Su and Koschny, Thomas and Zhou, Lin and Kramer, Matthew J. and Murthy, Akshay A. and Bal, Mustafa and Crisa, Francesco and Garattoni, Sabrina and Zhu, Shaojiang and Lunin, Andrei and Olaya, David and Hopkins, Peter and Romanenko, Alex and Grassellino, Anna and Wang, Jigang},
	title = {{Terahertz near-field imaging of sidewall-induced losses in superconducting qubits}},
	journal = {Appl. Phys. Lett.},
	volume = {127},
	number = {11},
	pages = {114002},
	year = {2025},
	month = sep,
	issn = {0003-6951},
	publisher = {AIP Publishing},
	doi = {10.1063/5.0284028}
}

@article{Bienfait2016Mar,
	author = {Bienfait, A. and Pla, J. J. and Kubo, Y. and Zhou, X. and Stern, M. and Lo, C. C. and Weis, C. D. and Schenkel, T. and Vion, D. and Esteve, D. and Morton, J. J. L. and Bertet, P.},
	title = {{Controlling spin relaxation with a cavity}},
	journal = {Nature},
	volume = {531},
	pages = {74--77},
	year = {2016},
	month = mar,
	issn = {1476-4687},
	publisher = {Nature Publishing Group},
	doi = {10.1038/nature16944}
}

@article{Yu2022Oct,
	author = {Yu, Liuqi and Matityahu, Shlomi and Rosen, Yaniv J. and Hung, Chih-Chiao and Maksymov, Andrii and Burin, Alexander L. and Schechter, Moshe and Osborn, Kevin D.},
	title = {{Experimentally revealing anomalously large dipoles in the dielectric of a quantum circuit}},
	journal = {Sci. Rep.},
	volume = {12},
	pages = {16960},
	year = {2022},
	month = oct,
	issn = {2045-2322},
	publisher = {Nature Publishing Group},
	doi = {10.1038/s41598-022-21256-7}
}

@article{Mann2025Aug,
	author = {Mann, Zachary and Cao, Ningping and Laflamme, Raymond and Zhou, Sisi},
	title = {{Quantum Error-Corrected Non-Markovian Metrology}},
	journal = {PRX Quantum},
	volume = {6},
	number = {3},
	pages = {030321},
	year = {2025},
	month = aug,
	publisher = {American Physical Society},
	doi = {10.1103/wfyl-wtz3}
}


\renewcommand{\theequation}{S.\arabic{equation}}
\setcounter{equation}{0}

\onecolumngrid
\newpage 

\section*{SUPPLEMENTAL MATERIAL}

\section{qubit parameters}
\label{SUP_qubit_params}
\subsection*{Device }

\begin{figure}[!htbp]
\includegraphics[width = .9\columnwidth]{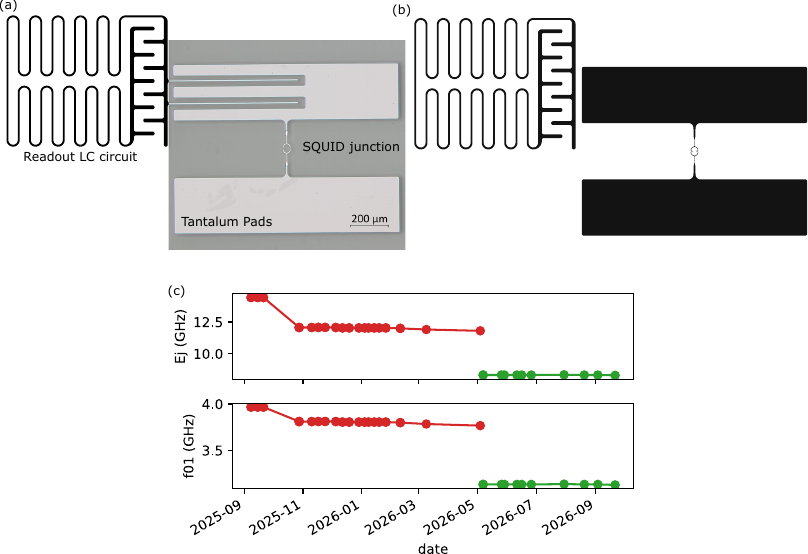}
\caption{ (a) Optical Image of the Transmon circuit of sample 1 next to scaled pattern of the readout LC structure. (b)Structure of sample 2 . It has the same exact structure without the two lines of the readout circuit extending into the transmon.  (b) Variation of the two measured transmons Josephson energies(red for sample 1 and green points for sample 2) over time. The sample 2 had already aged for a year before starting measurements.}
\label{fig_supp_parameter_changes}
\end{figure}

Both qubits follow the design shown in Fig.~\ref{fig_supp_parameter_changes}(a). They were fabricated on c-plane sapphire substrates, which were cleaned in isopropanol and heated to 400\,$^\circ$C in the ultra-high-vacuum deposition chamber overnight. A 200\,nm tantalum film was deposited by DC magnetron sputtering for 55~min at .003~mbar Ar pressure, at a substrate temperature of 800\,$^\circ$C. The capacitor pads and resonators were patterned by optical lithography (S1805 resist, Heidelberg $\mu$MLA maskless aligner) and etched by reactive-ion etching in an SF$_6$ plasma, preceded by a short O$_2$ descum. After resist stripping in hot DMSO, the wafer was dipped in 7:1 buffered oxide etch for 15\,min. The latter step is known to reduce loss in tantalum qubits, likely by thinning the native tantalum oxide surface layer as well as removing etch residues.

The Al/AlO$_x$/Al Josephson junctions forming the SQUID were defined by electron-beam lithography in an MMA(8.5)MAA/PMMA bilayer resist (EL13/A4) and deposited by double-angle shadow evaporation in a \emph{Plassys MEB550S2-I UHV} electron-beam evaporator. Prior to deposition, the exposed tantalum was cleaned in situ by Ar ion milling to ensure galvanic contact between the junction leads and the pads. The junctions consist of 25\,nm and 40\,nm Al layers separated by a static oxidation of 3.5\,min at 15\,mbar O$_2$. Lift-off was performed in warm acetone and N-ethyl-2-pyrrolidone (NEP), after which the wafer was diced.

The samples were measured over multiple cool downs and the values of $\mathrm{E_j}$ lowered through aging. The first qubit (red markers in Fig.~\ref{fig_supp_parameter_changes}(b)) was measured quickly after fabrication, which could explain the higher drop of Josephson energy after the first cool down.

\subsection*{Measurement setup}
\begin{figure}[!htbp]
\includegraphics[width = .7\columnwidth]{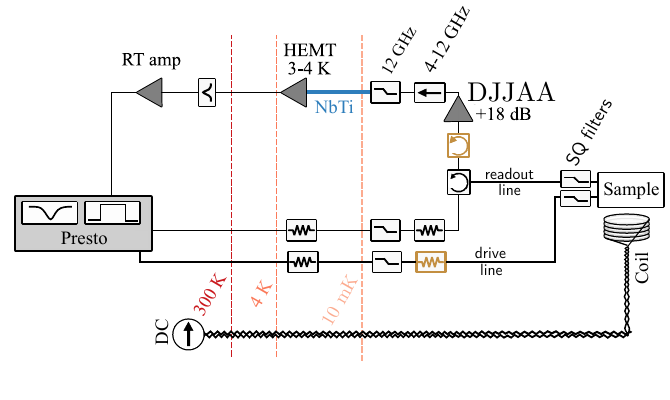}
\caption{ Setup of the experiment, the orange circulator was added as well as the orange attenuator changed from 20 dBm to 3 times 5dBm before the third cooldown shown in Table \ref{tab:coherence}. All the other attenuators are 20 dBm}
\label{fig_setup}
\end{figure}

The measurement setup is shown in Fig.~\ref{fig_setup}. The samples are measured in a copper waveguide setup similar to Ref.~\cite{Maleeva_2018}. 
The variations of $T_1$ and $T_2$ are summarized in Table~\ref{tab:coherence}.

\begin{table}[htbp]
  \centering
  \renewcommand{\arraystretch}{1.5}   
  \setlength{\tabcolsep}{12pt}        
  \caption{Measured coherence times and observed TLS across cooldowns.}
  \label{tab:coherence}
  \begin{tabular}{|c|l|c|c|c|}
    \hline
    \# & Date & $T_1$ (\si{\micro\second}) & $T_2$ ramsey (\si{\micro\second}) & TLS (\si{\giga\hertz}) \\
    \hline\hline
    1  & May 2025       & 20--25   & ---         & ---   \\ \hline
    2  & May 2025       & 65       & ---         & ---   \\ \hline
    3  & Jun--Sep 2025  & 125      & 10-20          & 3.948 \\ \hline
    4  & Nov 2025       & 105--120 & 50          & 3.819 \\ \hline
    5  & Early Dec 2025 & 100--120 & 20          & ---   \\ \hline
    6  & Dec 2025       & 110--120 & 80 $\to$ 20 & ---   \\ \hline
    7  & Jan 2026       & 120--130 & 30          & ---   \\ \hline
    8  & Mar 2026       & 120--150 & 20          & ---   \\ \hline
    9  & Apr--May 2026  & 80       & 20          & 3.044 \\ \hline
    10 & Jun 2026       & 80--100  & 20          & ---   \\ \hline
    11 & Jul 2026       & 50--70   & 15--20      & ---   \\ \hline
  \end{tabular}
\end{table}

\newpage

\section{Stark shift calibration}
\label{SUP_Stark_shift}

The Stark shift induced by a strong off-resonant drive follows, to leading order in the transmon's weak anharmonicity $\alpha$ based on the first two levels,
\begin{equation}
    \Delta f = \frac{1}{2\pi}\,[\frac{\Omega_s^2}{4(\Delta_s)}-\frac{\Omega_s^2}{4(\Delta_s+\alpha)}]= \frac{1}{2\pi}\,\frac{\alpha\,\Omega_s^2}{4\,\Delta_s(\Delta_s+\alpha)}
\end{equation}
where $\Omega_s$ is the Rabi rate of the Stark drive and $\Delta_s$ its detuning from $f_{01}$. This expression is derived from second-order perturbation theory on the lowest levels of the transmon and is valid as long as $\Omega_s \ll \Delta_s$, i.e. far from the single- and resonance conditions $\Delta_s \approx 0$ and $\Delta_s \approx -\alpha$.

\begin{figure}[!htbp]
\includegraphics[width = .6\columnwidth]{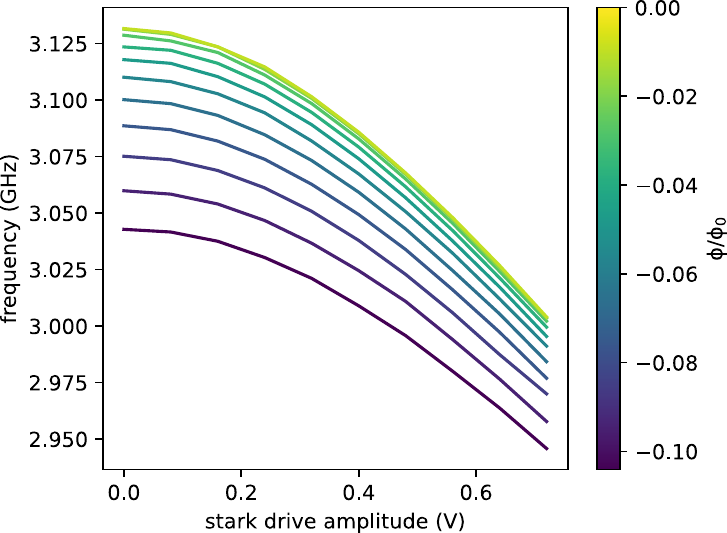}
\caption{ Calibration of the Stark shift performed for qubit 2 with a constant stark shift drive at 3.499 $\mathrm{GHz}$ in Volts defined at the Presto Output}
\label{fig_supp_calib_0}
\end{figure}

As $\Delta_s$ approaches the higher transitions of the qubit ladder ($\Delta_s \to -\alpha$, corresponding to the $1\!\to\!2$ transition), the leading-order expression above diverges and is no longer a good approximation. In practice we calibrate $\Delta f(\Omega_s, \Delta_s)$ directly via two tone at each working point rather than relying solely on the analytic formula. 
As shown in Fig.~\ref{fig_supp_calib_0} we can extract the frequency of the qubit for every flux and stark shift amplitude while using a constant Stark Drive at 3.499~GHz, which is 400 MHz above the transition frequency.

\begin{figure}[!htbp]
\includegraphics[width = 1\columnwidth]{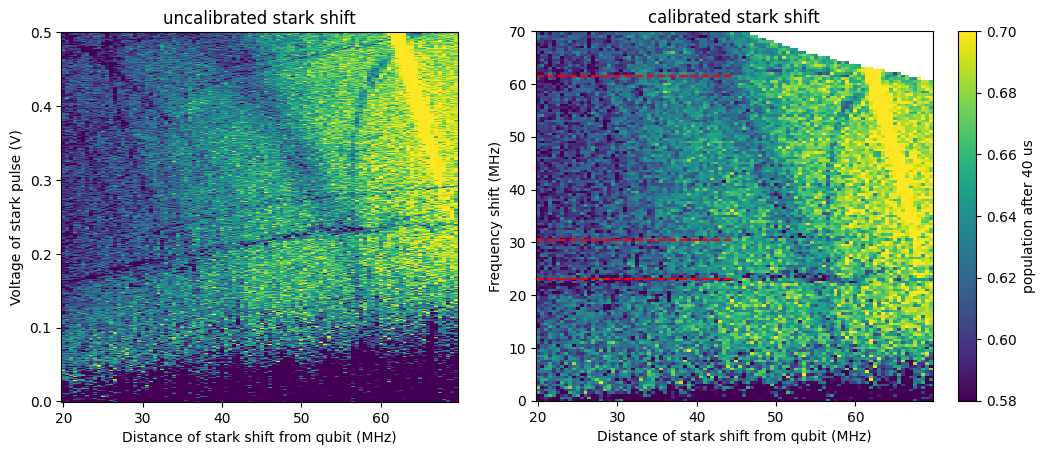}
\caption{ (a) Uncalibrated Data at varying Stark shift distance (b) Same data with calibration performed at each stark shift tone distance. We see many lines that are Stark shift tone dependent, indicating that they originate from non linear processes that include the stark shift tone.}
\label{fig_supp_calib_1}
\end{figure}

We also make use of closer Stark Shift drive on a case by case basis, calibrated in the same way. As shown in Fig.~\ref{fig_supp_calib_1} we can perform the same Quantum jump experiment using different Stark Shift drive detuning from the Transmon frequency. By calibrating each of them separately we readily differentiate lines that are not Stable with Stark Shift drive and thus are linked with a DUST process involving the drive.  

The first measurement with detuning performed on TLS 1 of Fig.~\ref{fig_transition_plot} was performed using 50 MHz detuning and the different detunings where performed by actively changing the qubit flux offset and recalibrating the experiment at every frequency. Subsequent measurement used the Stark shift during delay to sweep the detuning.

The stark shift amplitude going into the fridge depends by about 1 per thousand with the temperature of the room temperature amplifier that sends the pulse to the fridge. This requires the calibration to be made regularly, as well as prevents proper averaging of the signal for very small linewidths.

\newpage

\section{Coherence times and Full Quantum Jump signal close to TLS-qubit resonance}
\label{Supp_in_resonance}

For TLS1 in figure \ref{fig_transition_plot}, we can perform the Quantum jump experiment with a finer sweep at low detuning interleaved with $\mathrm{T_1}$ and $\mathrm{T_2}$ experiments by moving the qubit's flux. It shows that the decay experiment performed on a long-lived TLS with enough delay between the qubit resets still shows a lowering of the $\mathrm{T_1}$ when in resonance.  The decoherence time also shows the dephasing induced by the long lived TLS .

\begin{figure}[!htbp]
\includegraphics[width = .8\columnwidth]{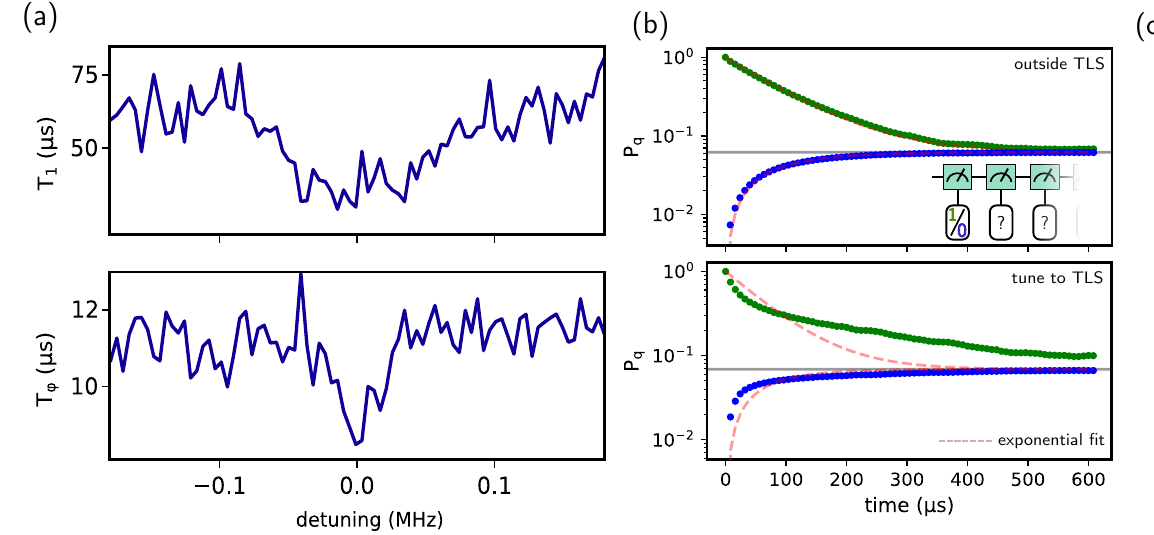}
\label{fig_supp_line}
\end{figure}

\newpage

\section{Quantum Jump spectroscopy for the other TLS}
\label{SUP_otherstark_shift}

The 3 TLS shown in figure \ref{fig_purcell} where found over two qubits and we exhibit the spectroscopy performed to find each of those TLS. 

\begin{figure}[!htbp]
\includegraphics[width = .6\columnwidth]{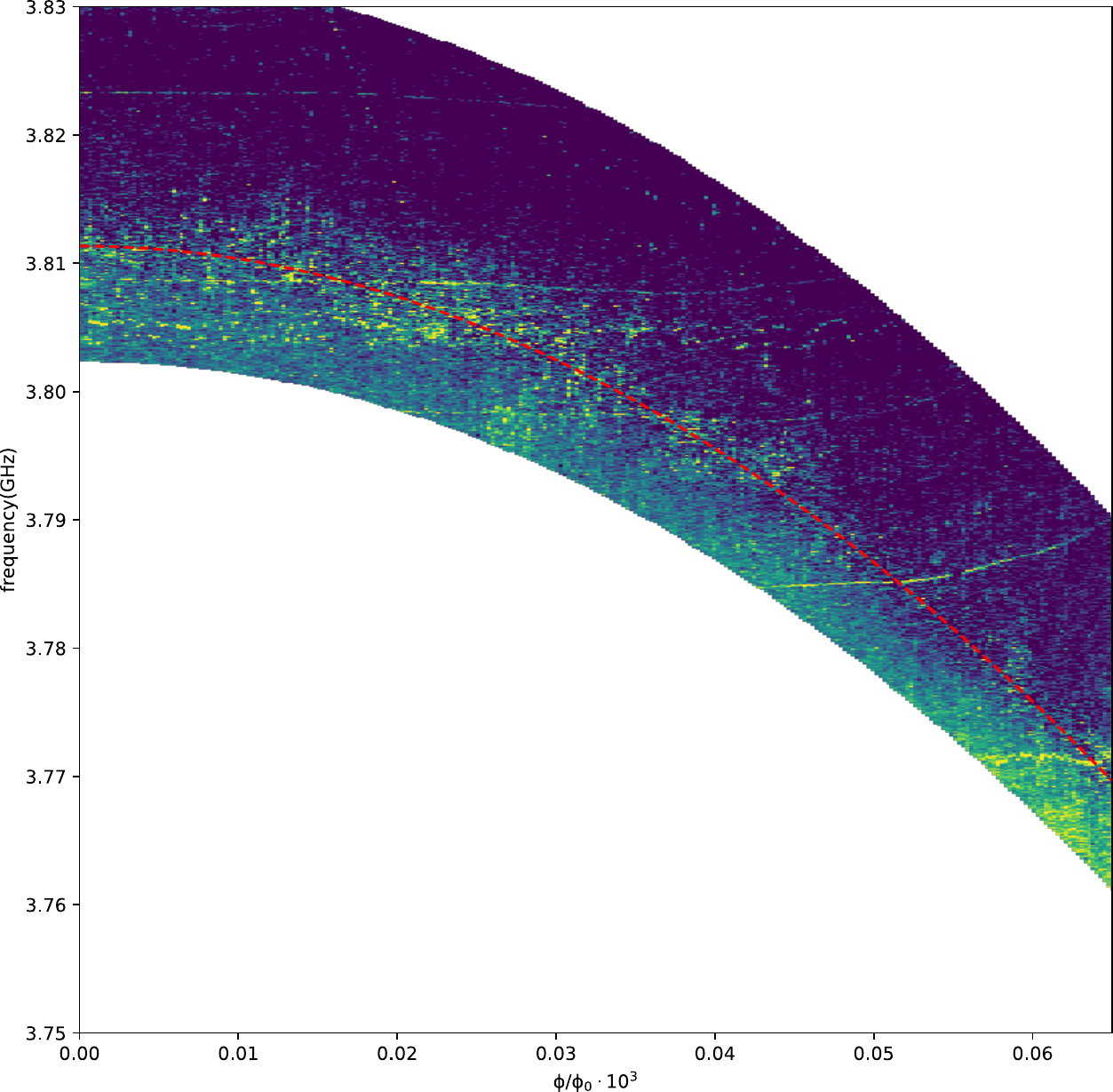}
\caption{ Quantum Jump spectroscopy of TLS2 that can be seen at 3.824 GHz. The calibration was only made up to flux .05 which explain the curvature of the lines for fluxes above this value. An extra MIST line can be seen around .02 of flux.}
\label{fig_supp_TLS2}
\end{figure}

The Spectroscopy allowing to find the TLS at 3.824 GHz is shown in Fig.~\ref{fig_supp_TLS2} and was performed in the same way as Fig.~\ref{fig_transition_plot}.

\newpage

\begin{figure}[!htbp]
\includegraphics[width = \columnwidth]{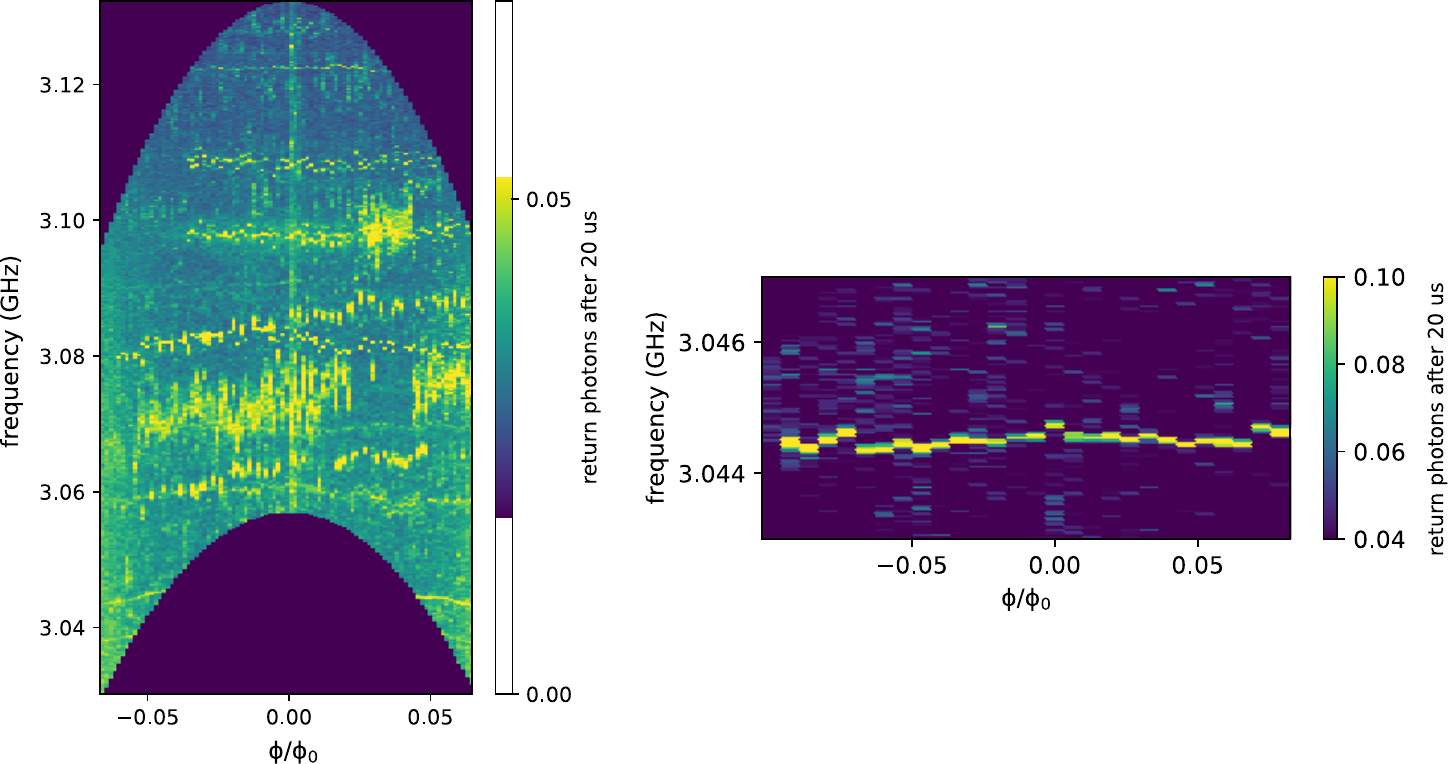}
\caption{ Quantum Jump spectroscopy of TLS3 performed close to zero field and zoom of the line measured at - 10 $\mathrm{\phi_0}$. The spectroscopy is performed only below the qubit. The spectroscopy took about 10 hours. The broad noisy TLS lines at 3.06, 3.07, 3.08 and 3.1~GHz were diffusing upward in frequency probably due to time evolution and subsequently disappeared.}
\label{fig_supp_TLS3}
\end{figure}

The Spectroscopy allowing to find the TLS at 3.044 GHz is shown in Fig.~\ref{fig_supp_TLS3} and was performed 
by using the technique found in Fig.~\ref{fig_zoom} and calibrating the stark shifts and Rabi pulses at every flux. This qubit had a temperature of 10 mK at 3.13~GHz giving around 1 \% of state 1, which rendered the Quantum jump technique unpractical. A close up of the TLS through more flux values is shown in Fig.~\ref{fig_supp_TLS3}(right) which shows the stability of the line in flux within the accuracy of the calibration.

\end{document}